&pdflatex

\documentclass[10pt,twocolumn,prl,aps,amssymb,amsmath,tightenlines,showpacs]{revtex4}

\usepackage{graphicx}

\newcommand{\half}{\textstyle{\frac{1}{2}}}

\newcommand{\cP}{{\cal P}}
\newcommand{\cT}{{\cal T}}

\newcommand{\cD}{{\cal D}}
\newcommand{\cH}{{\cal H}}
\newcommand{\cPT}{{\cal PT}}

\begin{document}


\title{Infinitely many inequivalent field theories from one Lagrangian}

\author{Carl~M.~Bender$^{1,2,3}$,\email{cmb@wustl.edu}
Daniel~W.~Hook$^{1,4}$,\email{d.hook@imperial.ac.uk}
Nick~E.~Mavromatos$^{2,5}$,\email{nikolaos.mavromatos@kcl.ac.uk}
and Sarben Sarkar$^2$\email{sarben.sarkar@kcl.ac.uk}}

\affiliation{$^1$Department of Physics, Washington University, St.~Louis, MO
63130, USA\\
$^2$Department of Physics, King's~College~London, London WC2R 2LS, UK\\
$^3$Department of Mathematical Science, City University London,
Northampton Square, London EC1V 0HB, UK\\
$^4$Department of Physics, Imperial College London, London SW7 2AZ, UK\\
$^5$Theory Division, CERN, CH-1211 Geneva 23, Switzerland}

\date{\today}

\begin{abstract}
Logarithmic time-like Liouville quantum field theory has a generalized $\cPT$
invariance, where $\cT$ is the time-reversal operator and $\cP$ stands for an
$S$-duality reflection of the Liouville field $\phi$. In Euclidean space the
Lagrangian of such a theory, $L=\half(\nabla\phi)^2-ig\phi\exp(ia\phi)$, is
analyzed using the techniques of $\cPT$-symmetric quantum theory. It is shown
that $L$ defines an infinite number of unitarily inequivalent sectors of the
theory labeled by the integer $n$. In one-dimensional space (quantum mechanics)
the energy spectrum is calculated in the semiclassical limit and the $m$th
energy level in the $n$th sector is given by $E_{m,n}\sim(m+1/2)^2a^2/(16n^2)$.
\end{abstract}


\maketitle

Motivated by studies of time-like logarithmic Liouville quantum field theory, we
examine here the interaction $-ig\phi\exp(ia\phi)$ in field theory and its
quantum-mechanical analog $-igx\exp(iax)$. This remarkable interaction gives
rise to a countably infinite number of inequivalent quantum theories.

The interaction $-ig\phi\exp(ia\phi)$ has its origin in conformal field theory
(CFT) of Liouville type, whose interaction has the form $e^{\alpha\phi}$
\cite{R1,R2,R3,R4,R5,R6}. This exponential arises in string theory and in
two-dimensional gravity, which are defined on two-dimensional manifolds. Using
general coordinate invariance, one can show that the metric tensor $g_{\mu\nu}$
for these theories can be reduced locally to $g_{\mu\nu}=\eta_{\mu\nu}e^{\alpha
\phi}$, where $\eta_{\mu\nu}$ is the Minkowski metric. String theory and
two-dimensional gravity are conformally invariant at the classical level, but
quantum effects can produce an anomaly that destroys conformal invariance.
Conformal symmetry is restored if the field $\phi$ is governed by the Liouville
action
$$S=\int_{-\infty}^\infty d\tau\int_0^{2\pi} d\sigma\left[\half
(\partial_\tau\phi)^2-\half(\partial_\sigma\phi)^2-g e^{\alpha\phi}\right].$$

Recoil effects for zero-dimensional D-branes scattering off closed strings are
described by the interaction $\phi e^{\alpha\phi}$ in addition to the usual
Liouville interaction $e^{\alpha\phi}$ \cite{R7}. Such pairs of operators
define a logarithmic CFT \cite{R8}. Logarithmic CFTs also arise in descriptions
of quenched disordered condensed matter systems \cite{R9,R10}. Supercritical
strings \cite{R11} and the condensation of tachyons \cite{R12,R13} is studied
in the context of {\it time-like} Liouville theories, whose interaction term has
$\alpha$ replaced by $ia$. Thus, combining the ideas of Liouville and
logarithmic CFT, we are led to consider the $d$-dimensional Euclidean
Lagrangian
\begin{equation}
L=\half(\nabla\phi)^2-ig\phi e^{ia\phi}-he^{ia\phi},
\label{e1}
\end{equation}
where $\phi$ is a scalar field and $a$, $g$, and $h$ are treated as positive
real parameters.

The Lagrangian (\ref{e1}) is not Hermitian and one cannot make such a theory
Hermitian by adding its Hermitian conjugate because this would destroy the
conformality property of the theory. Nevertheless, the techniques of $\cPT$
quantum theory \cite{R14} can be used to study this field theory. The Lagrangian
is not obviously $\cPT$ invariant because in Liouville theory the field $\phi$
is assumed to transform as a scalar, so it does not change sign under space
reflection. [If $\phi$ were a pseudoscalar field, the Lagrangian would be $\cPT$
invariant because under parity reflection $\cP$, $\phi$ would change sign $\cP
\phi(x,t)\cP=-\phi(-x,t)$, and under time reversal $\cT$, $i$ changes sign $\cT
i\cT=-i$.] However, we let $\cP$ represent an {\it S duality} reflection
\cite{R15},
\begin{equation}
\cP\phi(x,t)\cP=-\phi(x,t),
\label{e2}
\end{equation}
and with this definition of $\cP$, $L$ is manifestly $\cPT$ symmetric. A
non-Hermitian $\cPT$-invariant theory can have a positive real spectrum
and unitary time evolution \cite{R14}.

The interaction terms of (\ref{e1}) have a periodic component and thus $L$ bears
a resemblance to some previously studied $\cPT$-symmetric theories, including
the complex Toda lattice \cite{R16}, complex diffraction gratings \cite{R17},
and complex crystal lattices \cite{R18}. Complex $\cPT$-symmetric periodic
potentials exhibit real-energy band structure. There have also been studies of
the complex sine-Gordon equation \cite{R19}, complex dynamical systems
\cite{R20}, and $\cPT$-symmetric exponential potentials \cite{R21}. However, the
factor of $\phi$ multiplying $g$ in (\ref{e1}), which is characteristic of
logarithmic CFT, leads to surprising new effects. Specifically, the partition
function as a path integral over $L$,
\begin{equation}
Z_n=\int\cD\phi\exp\left(\int d^dx\,L\right),
\label{e3}
\end{equation}
has {\it infinitely} many distinct functional integration paths labeled by
$n=1,\,2,\,3,\,\ldots$, each defining a valid but unitarily {\it inequivalent}
quantum theory. This multiplicity of theories is not due to monodromy (there is
no winding number because the integrand is entire) nor is it a topological
effect (like $\theta$ vacua). The quantum-mechanical version of this time-like
logarithmic CFT has discrete energy levels rather than energy bands. The $m$th
energy level in the $n$th theory grows like $m^2$ as $m\to\infty$, but for fixed
$m$ the energies decay like $n^{-2}$ as $n\to\infty$.

To find the integration paths on which the integral (\ref{e3}) converges, we
must locate in field space the pairs of {\it Stokes wedges} inside which the
integrand vanishes exponentially. To begin, we simplify this integral by
shifting $\phi$ by a constant to eliminate the parameter $h$ in $L$. Next, we
neglect the effect of the kinetic term $(\nabla\phi)^2$ because it does not
affect the convergence. We also ignore the spatial integral in the exponent and
study the convergence at each lattice point separately; that is, we perform an
{\it ultralocal} analysis \cite{R22} and examine the convergence of
\begin{equation}
I=\int d\phi\,\exp\left(ig\phi e^{ia\phi}\right).
\label{e4}
\end{equation}

\begin{figure}[h!]
\begin{center}
\includegraphics[scale=0.31]{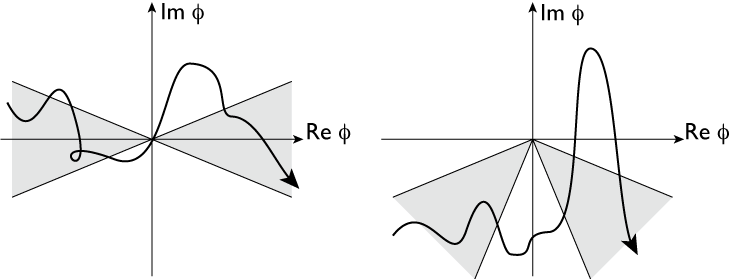}
\end{center}
\caption{Integration paths terminating in Stokes wedges (shaded regions) for
Hermitian $\phi^4$ (left panel) and $\cPT$-symmetric $-\phi^4$ (right panel)
interactions.}
\label{F1}
\end{figure}

We illustrate how to locate Stokes wedges in the complex-$\phi$ plane by using
monomial potentials $\phi^k$. For such potentials the angular opening of
the Stokes wedges has a simple $k$ dependence. For $k=4$ the integral $\int d
\phi\,\exp\left(-\phi^4\right)$ converges in a pair of Stokes wedges of angular
opening $45^\circ$ centered about the positive-$\phi$ and negative-$\phi$ axes
(Fig.~\ref{F1}, left panel). The integration contour must terminate inside these
Stokes wedges. For a $\cPT$-symmetric upside-down $-\phi^4$ potential, the
associated integral $\int d\phi\,\exp\left(\phi^4\right)$ converges in a pair of
Stokes wedges of angular opening $45^\circ$ centered about ${\rm arg}\,\phi=-
45^\circ$ and ${\rm arg}\,\phi=-135^\circ$ (Fig.~\ref{F1}, right panel). 

The integral $I$ in (\ref{e4}) is unusual because it converges in pairs of
Stokes wedges of asymptotic angular opening $0^\circ$. There are {\it
infinitely} many such Stokes wedges in the complex-$\phi$ plane, all parallel to
the negative imaginary axis. Each pair of wedges defines a distinct physical
theory having its own real energy spectrum. Guralnik {\it et al} \cite{R23}
first recognized that for functional integrals, inequivalent classes of contours
with different complex boundary conditions are associated with nonunique
solutions to the Dyson-Schwinger equations. They found that multiple solutions
account for inequivalent $\theta$ vacua \cite{R23}. In Ref.~\cite{R24} it was
shown that if the pairs of Stokes wedges possess left-right symmetry ($\cPT$
symmetry) in complex field space, the field theory is physically acceptable
because the masses (poles of the Green's functions) are real and the theory is
unitary. However, Ref.~\cite{R24} only considered the case of a {\it finite}
number of distinct physical theories, one theory for each pair of wedges.

\begin{figure}[h!]
\begin{center}
\includegraphics[scale=0.43]{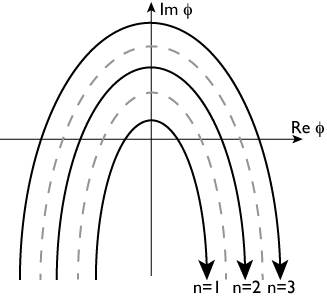}
\end{center}
\caption{Integration paths in complex-$\phi$ space for which the integral
(\ref{e4}) converges. The paths terminate in infinitely thin Stokes wedges and
define unitarily inequivalent theories.}
\label{F2}
\end{figure}

Here, we consider the unusual case of an infinite number of inequivalent
theories corresponding to pairs of infinitely thin Stokes wedges. To find the
paths of integration on which $I$ converges, we introduce polar coordinates
$\phi=Re^{i\theta}$ and treat $R$ as large. Then, (\ref{e4}) becomes
\begin{equation}
I=\int d\phi\,\exp\left(igRe^{-aR\sin\theta}e^{i\theta+iaR\cos\theta}\right).
\label{e5}
\end{equation}
We need to find Stokes wedges, that is, the angles at which the integrand
vanishes exponentially fast as $R\to\infty$. At the center $\theta$ of a Stokes
wedge, the exponent in (\ref{e5}) must be {\it real} and thus $\theta+aR\cos
\theta\sim(n+1/2)\pi$ as $R\to\infty$ ($n$ integer). Also, the argument of the
exponential must be {\it negative} so that it vanishes as $R\to\infty$. Thus,
$\sin(\theta+aR \cos\theta)\sim1$ as $R\to\infty$ and we find that
\begin{equation}
\theta+aR\cos\theta\sim(\pm 2n+1/2)\pi\quad(R\to\infty),
\label{e6}
\end{equation}
where $n>0$. The {\it maximum} rate of decay occurs at the center of the wedge,
so $\theta$ must be close to $-\pi/2$. Hence, we substitute $\theta=-\pi/2+
\epsilon$ into (\ref{e6}) and obtain $\epsilon\sim(2n+1)\pi/(aR)$ $(R\to\infty
)$. We find that the centers of the Stokes wedges lie at
\begin{equation}
\theta_n\sim-\half\pi\pm\frac{(2n+1)\pi}{aR}\quad(R\to\infty).
\label{e7}
\end{equation}
To summarize, for the partition function $Z_n$ in (\ref{e3}) the $n$th path of
functional integration originates in the $-n$th Stokes wedge, terminates in the
$n$th Stokes wedge, and is asymptotically parallel to the negative-imaginary
axis. The path is $\cPT$ (left-right) symmetric (see Fig.~\ref{F2}).

The $\cPT$-symmetric quantum-mechanical Hamiltonian $\cH$ corresponding to the
field-theoretic Lagrangian (\ref{e1}) is
\begin{equation}
\cH=p^2-igx e^{iax}+he^{iax},
\label{e8}
\end{equation}
where $a$, $g$, and $h$ are assumed to be real and positive. As we did for the
the field-theoretic model, we shift $x$ by a constant to eliminate $h$ and
obtain the Hamiltonian
\begin{equation}
\cH=p^2-igx e^{iax}.
\label{e9}
\end{equation}

\begin{figure}[h!]
\begin{center}
\includegraphics[scale=0.12]{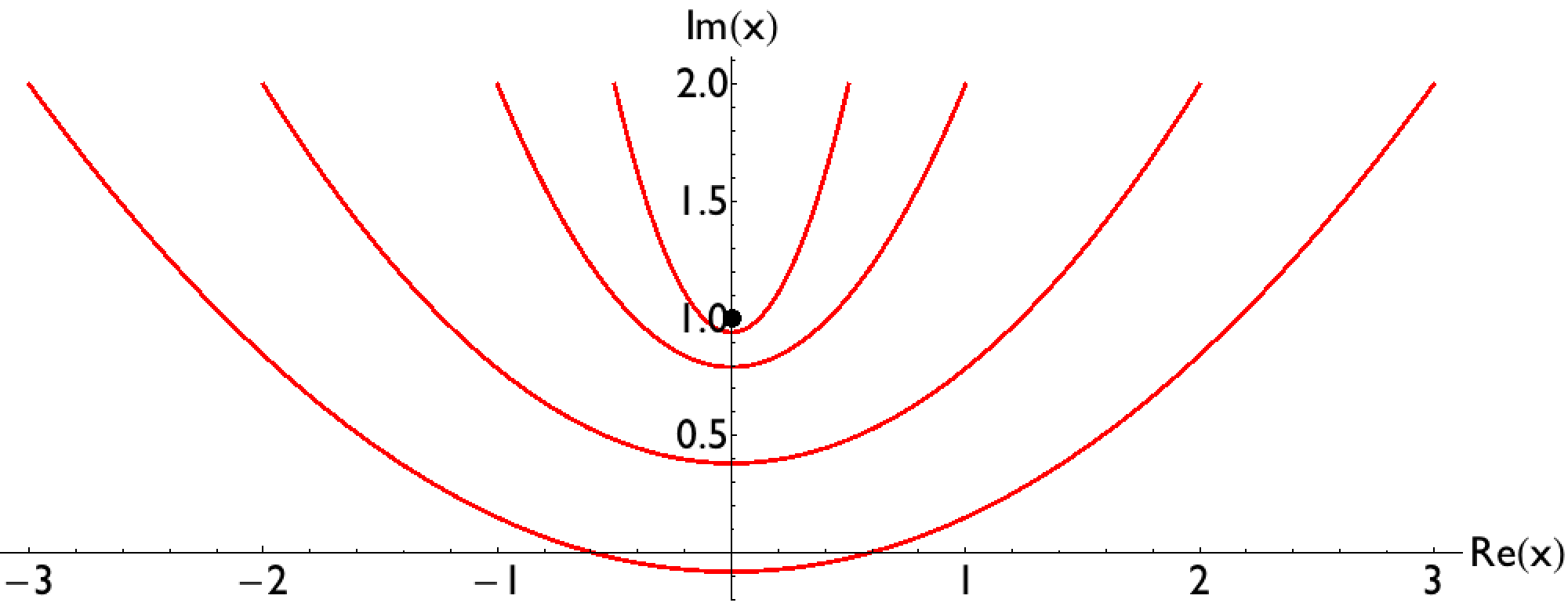}
\end{center}
\caption{Complex classical paths for the potential $V=-ix$ with $E=1$. The paths
are parabolic and thus do not close.}
\label{F3}
\end{figure}

Both quantum-mechanical Hamiltonians (\ref{e8}) and (\ref{e9}) possess a
singular limit. If $a\to0$, $\cH$ in (\ref{e9}) reduces to $\cH=p^2-igx$. This
limit is singular because, as was shown by Herbst, the spectrum of this
Hamiltonian is null \cite{R25}. To explain intuitively the absence of
eigenvalues, we solve Hamilton's classical equations $\dot{x}=\partial\cH/
\partial p=2p$, $\quad\dot{p}=-\partial\cH/\partial x=ig$. Combining these
equations gives $\ddot{x}=2ig$, whose solutions are parabolas in the complex
plane: $x(t)=igt^2+\alpha t+\beta$ ($\alpha$ and $\beta$ are constants.)
Parabolas are {\it open} curves (see Fig.~\ref{F3}), so it is not possible to
satisfy the Bohr-Sommerfeld (WKB) quantization condition, $\oint dx\sqrt{E-V(x)}
=(m+1/2)\pi$ $(m=0,\,1,\,2,\,\ldots)$, which involves an integral over a {\it
closed} path.

\begin{figure}[t!]
\begin{center}
\includegraphics[scale=0.12]{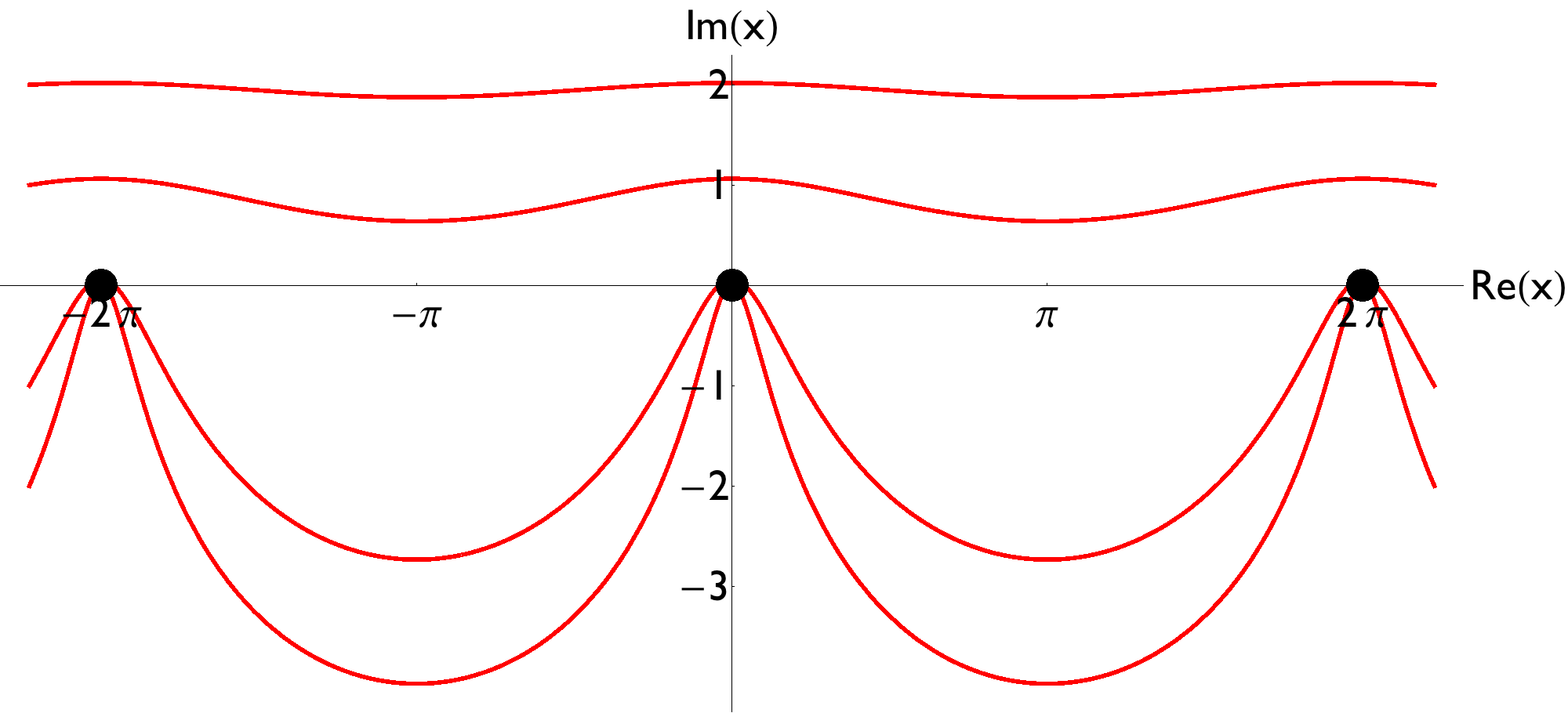}
\end{center}
\caption{Classical paths for $H=\half p^2+e^{ix}$ at $E=1$. The classical paths
are $2\pi$-periodic and open.}
\label{F4}
\end{figure}

A singular limit of $\cH$ in (\ref{e8}) is $g\to0$, which gives the Hamiltonian
$\cH=p^2+he^{iax}$ studied in Refs.~\cite{R17,R18}. This Hamiltonian exhibits
real energy bands, but has no discrete energies. To explain intuitively the
absence of discrete eigenvalues we plot the classical paths in the complex plane
(see Fig.~\ref{F4}). We see that these classical paths are $2\pi$-periodic and
thus are open curves.

\begin{figure}[t!]
\begin{center}
\includegraphics[scale=0.12]{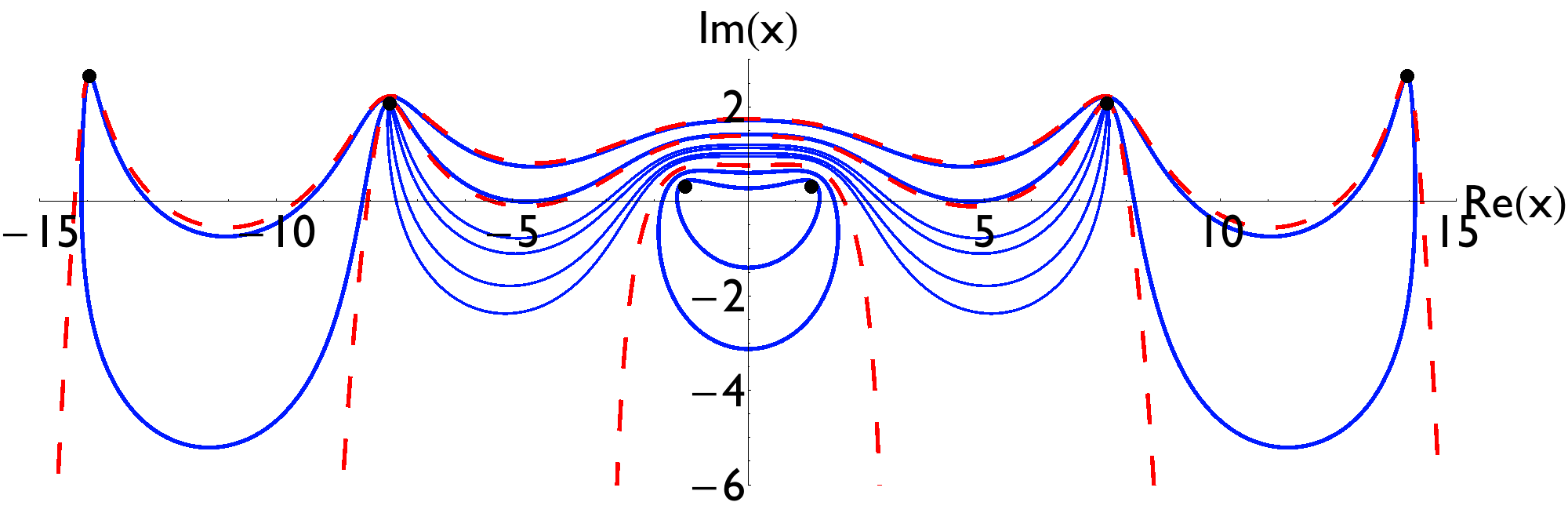}
\end{center}
\caption{Classical paths (solid curves) and separatrices (dashed curves) for
$\cH$ in (\ref{e9}) for $a=1$, $g=1$, and $E=1$. The central paths enclose the
$n=\pm1$ pair of turning points (dots); the next family encloses the $n=\pm2$
pair, and so on.}
\label{F7}
\end{figure}

In contrast, the classical paths for $\cH$ in (\ref{e9}) are {\it closed}. (See
Fig.~\ref{F7}.) Thus, that Hamiltonian has discrete bound states. To prepare for
calculating the eigenvalues of the bound states of $\cH$ in (\ref{e9}) we must
locate the classical turning points, which are the roots of $E=-igx e^{iax}$. We
assume that $E$, $g$, and $a$ are all positive and let $x=A+iB$. Thus, we must
solve the transcendental equation
\begin{equation}
E=-ig(A+iB)e^{ia(A+iB)}.
\label{e10}
\end{equation}
The imaginary part of (\ref{e10}) gives $B$ in terms of $A$, $B=A\,{\rm cot}(a
A)$, and substituting this result into (\ref{e10}) gives $E\sin(aA)/(gA)=e^{-aA
{\rm cot}(aA)}$. So, if we let $\nu=aE/g$ and $\alpha=aA$, we get the
transcendental equation
\begin{equation}
\nu\alpha^{-1}\sin\alpha=e^{-\alpha\,{\rm cot}\alpha}.
\label{e11}
\end{equation}
To solve this equation graphically, we plot the left side of (\ref{e11}) as a
solid curve and the right side as a dotted curve. (The left and right sides are
even functions of $x$.) The intersections of these curves are solutions to
(\ref{e11}).

\begin{figure}[h!]
\begin{center}
\includegraphics[scale=0.12]{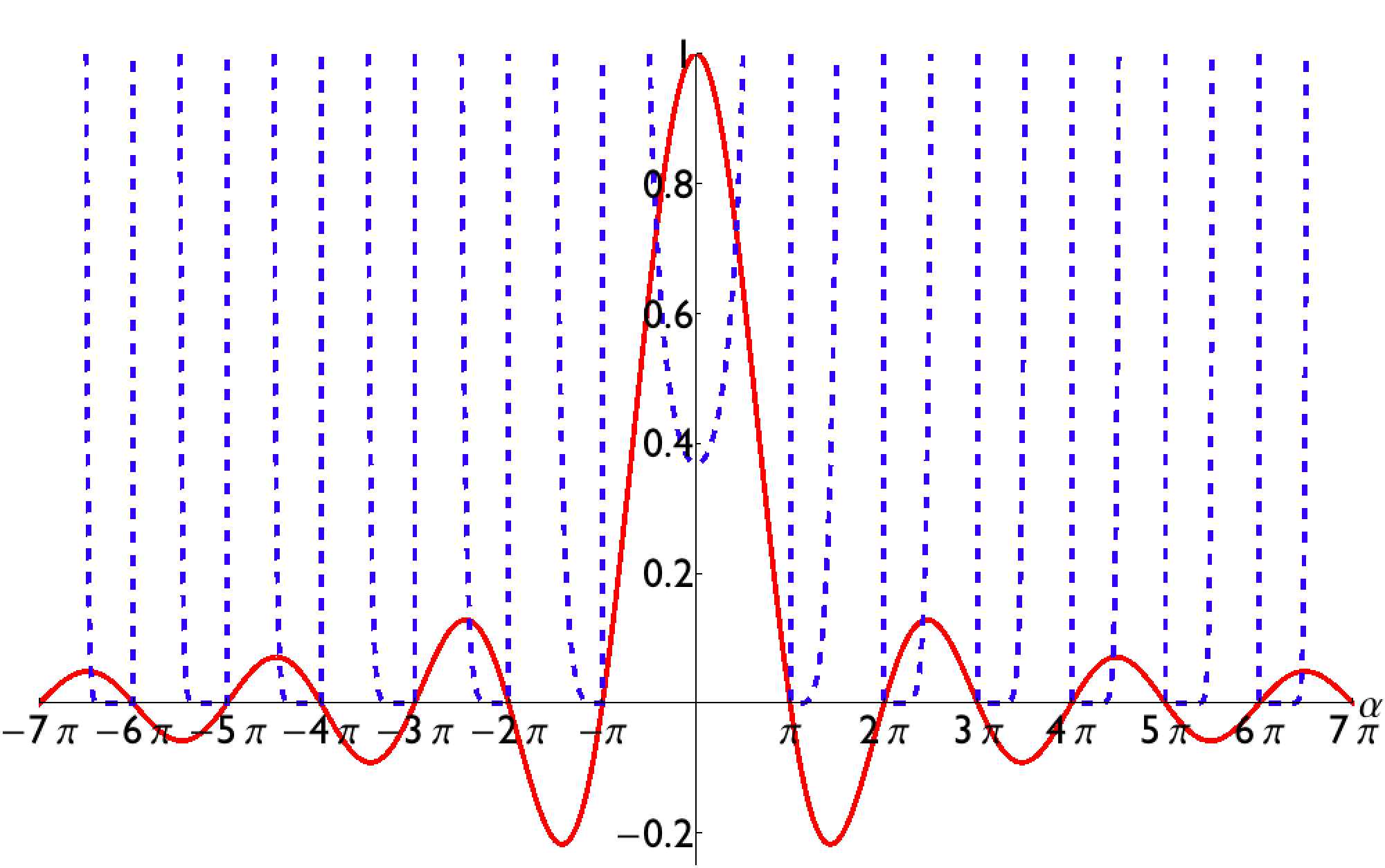}
\end{center}
\caption{Graphical solution to (\ref{e11}). The solid curve is the left side of
(\ref{e11}) and the dashed curve is the right side.}
\label{F5}
\end{figure}

Figure~\ref{F5} shows that there are two sets of solutions. The first is exactly
$x=n\pi$, but we reject this solution because it gives $B=\infty$. The second
set has intersection points near $\alpha_n=(2n+1/2)\pi-\delta$, where $\delta\ll
1$ as $n\to\infty$. Thus, for large $n$ the turning points are located
symmetrically about the imaginary-$x$ axis at
\begin{eqnarray}
{\rm Re}\,x&=&A\sim\pm(2n+1/2)\pi/a,\nonumber\\
{\rm Im}\,x&=&B\sim\frac{1}{a}\log\left[\frac{(2n+1/2)\pi g}{aE}\right].
\label{e12}
\end{eqnarray}
Table \ref{t1} verifies the accuracy of this asymptotic formula.

\begin{table}[h]
\begin{center}
\begin{tabular}{|c|c|c|}\hline
Turning point number & Exact & Approximate \\\hline
$n=0$ & $1.3372 + 0.3181i$ & $1.5708 + 0.4516i$ \\
$n=1$ & $7.5886 + 2.0623i$ & $7.85398 + 2.0610i$ \\
$n=2$ & $13.9492 + 2.6532i$ & $14.1372 + 2.6488i$ \\
$n=3$ & $20.2725 + 3.0202i$ & $20.4204 + 3.0165i$ \\
$n=4$ & $26.5805 + 3.2878i$ & $26.7035 + 3.2848i$ \\
\hline
\end{tabular}
\end{center}
\caption{Good agreement between the numerically precise values of the turning
points and the asymptotic approximation in (\ref{e12}) when $a=1$, $g=1$, and
$E=1$.}
\label{t1}
\end{table}

\begin{figure}[h!]
\begin{center}
\includegraphics[scale=0.12]{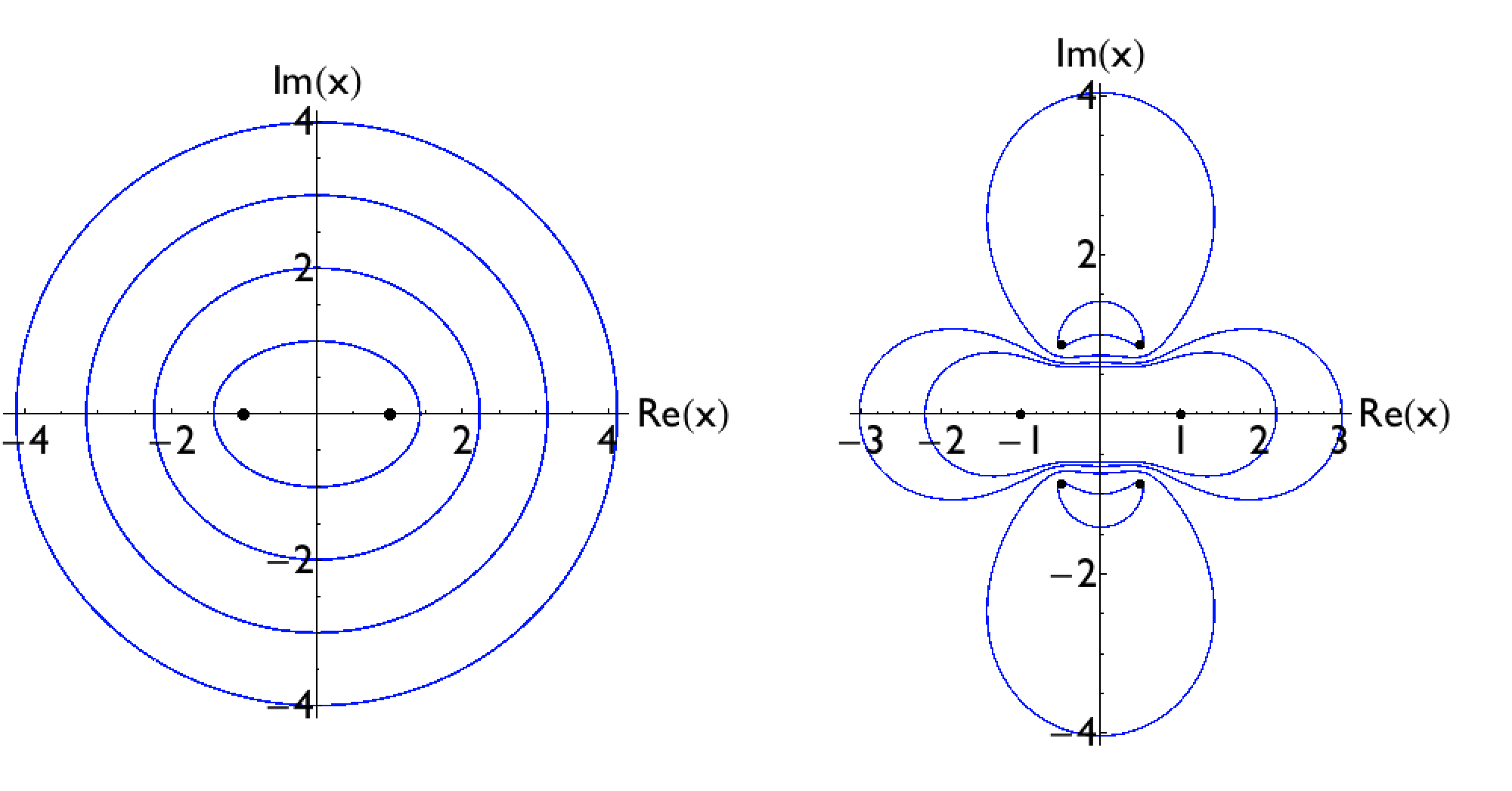}
\end{center}
\caption{Left panel: Four complex classical paths for the $p^2+x^2$ oscillator
with energy $E=1$. The family of paths encloses the turning points (dots) at
$x=\pm1$ and fills the entire complex plane. Right panel: Two examples of each
of the three families of classical paths for the $p^2+x^6$ oscillator for $E=1$.
Each family encloses one pair of turning points and together the three families
fill the complex plane.}
\label{F6}
\end{figure}

Having found the turning points, we next examine the complex classical paths for
$\cH$ in (\ref{e9}). In general, for any given Hamiltonian, the classical energy
determines a continuous family of classical paths distinguished by the initial
value of $x$. See, for example, the classical paths for the $x^2$ oscillator
in Fig.~\ref{F6} (left panel) and for the $x^6$ oscillator in Fig.~\ref{F6}
(right panel). Note that every family of classical paths encloses one pair of
turning points.

The classical paths for the Hamiltonian (\ref{e9}) are shown in Fig.~\ref{F7}.
The turning points are enclosed in pairs. Also shown are the separatrix paths
(see Table~\ref{t2}) that divide the families of closed classical trajectories.

\begin{table}[h]
\begin{center}
\begin{tabular}{|c|c|}\hline
Number of separatrix & Crossing point on the $y$ axis \\\hline
$n=1$ & $0.7613 i$ \\
$n=2$ & $1.3867 i$ \\
$n=3$ & $1.7485 i$ \\
\hline
\end{tabular}
\end{center}
\caption{Points where the separatrix paths in Fig.~\ref{F7} cross the imaginary
axis \cite{R26}.}
\label{t2}
\end{table}

Finally, we perform a leading-order WKB calculation of eigenvalues of the
Hamiltonian in (\ref{e9}). In general, when a theory is quantized, each pair of
classical turning points (and its continuous family of closed complex classical
paths) corresponds to a different and unitarily inequivalent quantum theory. For
example, the central pair of turning points in Fig.~\ref{F6} (right panel)
corresponds to the conventional Hermitian $x^6$ quantum oscillator, whereas the
upper and lower pairs correspond to $\cPT$-symmetric $x^6$ oscillators.
Thus, since there are an infinite number of pairs of classical turning points 
for the model in Fig.~\ref{F7}, we anticipate that there will be an infinite
number of classes of eigenvalues for the time-independent
Schr\"odinger equation for $\cH$ in (\ref{e9}),
\begin{equation}
-\psi''(x)-igxe^{iax}\psi(x)=E\psi(x).
\label{e13}
\end{equation}

The complex-$x$ WKB quantization condition is
\begin{equation}
\int_{x_{-n}}^{x_n} dx\,\sqrt{E+igxe^{iax}}=(m+1/2)\pi,
\label{e14}
\end{equation}
where the turning points $x_{\pm n}=\frac{1}{a}\left[\pm 2n\pi+i\log\left(\frac{
2n\pi g}{aE}\right)\right]$ are given in (\ref{e12}). For large $n$, (\ref{e14})
simplifies to
$$\frac{(m+1/2)a}{2n\sqrt{E}}\sim \int_{-1}^1 dw\,\sqrt{1+iwe^{2in\pi w}}
\sim 2\quad(n\to\infty).$$
Thus, for  large $n$ and large $m$, the WKB approximation to the $m$th energy in
the $n$th eigenspectrum is
\begin{equation}
E_{m,n}\sim(m+1/2)^2a^2n^{-2}/16.
\label{e15}
\end{equation}
The $m$th eigenvalue in the $n$th spectrum grows like the energies in a square
well \cite{R27}. However, the $m$th eigenvalue in the $n$th spectrum behaves
like the energy levels in the Balmer series for the hydrogen atom, and decays
like $n^{-2}$. The parameter $g$ does not appear in (\ref{e15}); it appears only
in higher-order WKB. The theories corresponding to different values of $n$ are
all inequivalent --- they are associated with different pairs of Stokes wedges
and have different energy spectra. To conclude, $\cPT$ analysis reveals the
simple but astonishing structure that underlies time-like Liouville logarithmic
CFT.

In future work we will investigate (i) whether it is possible to tunnel between
the inequivalent theories that we have found, and (ii) whether it is possible to
examine the conformal nature of two-dimensional logarithmic CFT by studying the
WKB approximation to the quantum-mechanical eigenfunctions in (\ref{e13})
\cite{R28}.

CMB was supported in part by the U.S. Department of Energy and NEM was supported
in part by the London Centre for Terauniverse Studies (LCTS), using funding from
the European Research Council via the Advanced Investigator Grant 267352 and by
STFC (UK) under the research grant ST/J002798/1. CMB and SS are grateful for a
Royal Society (UK) International Exchange grant.

\end{document}